\documentclass[final, 3p, times, compress]{elsarticle}

\usepackage{amsmath, amssymb, mathtools}
\usepackage{bm} 
\usepackage{amsfonts}

\usepackage{graphicx, subcaption} 
\usepackage{siunitx} 
\usepackage{tabularx}

\usepackage{ifthen} 

\usepackage{calc}
\usepackage[normalem]{ulem}
\newcommand{\soutmath}[1]{\ifmmode\text{\sout{\ensuremath{#1}}}\else\sout{#1}\fi}

\newcommand{\complexi}{\hspace{.15ex}\textrm{i}\hspace{.15ex}}

\newcommand{\calH}{\mathcal{H}}
\newcommand{\calL}{\mathcal{L}}

\newcommand{\xx}{\bm{x}}
\newcommand{\rr}{\bm{r}}

\newcommand{\AdS}{\mathrm{AdS}}
\newcommand{\SNG}{\ensuremath{S_\mathrm{N.G.}^{}}}
\newcommand{\zmin}{\ensuremath{z_\mathrm{min}}}
\newcommand{\zminbar}{\ensuremath{\bar{z}_\mathrm{min}}}

\newcommand{\sqrtsum}[2]{\sum\!\raisebox{-0.5ex}{$\rule{0ex}{2.3ex}_{\,#1}^{#2}$}}

\newlength\bracketsize

\newcommand{\PAR}[2][-1]{
    \setlength{\bracketsize}{0.75ex * #1 + 0.50ex}
    \ifthenelse
    { \equal{#1}{-1} }
    { \ensuremath{ {\left( #2 \right) } } }
    { \ensuremath{
            {               \left( \rule{0pt}{ \bracketsize } \right. \hspace{-1pt} }
            #2
            { \hspace{-1pt} \left. \rule{0pt}{ \bracketsize } \right)               }
} } }

\newcommand{\COL}[2][-1]{
    \setlength{\bracketsize}{0.75ex * #1 + 0.50ex}
    \ifthenelse
    { \equal{#1}{-1} }
    { \ensuremath{ {\left[\, #2 \,\right] } } }
    { \ensuremath{
            {               \left[ \rule{0pt}{ \bracketsize } \right. \hspace{-1pt} }
            \,#2\,
            { \hspace{-1pt} \left. \rule{0pt}{ \bracketsize } \right]               }
} } }

\newcommand{\CHA}[2][-1]{
    \setlength{\bracketsize}{0.75ex * #1 + 0.50ex}
    \ifthenelse
    { \equal{#1}{-1} }
    { \ensuremath{ {\left\{ #2 \right\} } } }
    { \ensuremath{
            {              \left\{ \rule{0pt}{ \bracketsize} \right.  \hspace{-1pt} }
            #2
            { \hspace{-1pt} \left. \rule{0pt}at{ \bracketsize} \right\}               }
} } }

\newcommand{\at}[2][-1]{
    \setlength{\bracketsize}{0.75ex * #1 + 0.50ex}
    \ifthenelse
    { \equal{#1}{-1} }
    { \ensuremath{ {\left.\, #2 \right| } } }
    { \ensuremath{
            {           \!\!\left. \rule{0pt}{ \bracketsize} \right. \hspace{-1pt} }
            \,#2
            { \hspace{-1pt} \left. \rule{0pt}{ \bracketsize} \right|              }
} } }

\newcommand{\abs}[2][-1]{
    \setlength{\bracketsize}{0.75ex * #1 + 0.50ex}
    \ifthenelse
    { \equal{#1}{-1} }
    { \ensuremath{ {\left| #2 \right| } } }
    { \ensuremath{
            {               \left| \rule{0pt}{ \bracketsize} \right. \hspace{-1pt} }
            #2
            { \hspace{-1pt} \left. \rule{0pt}{ \bracketsize} \right|               }
} } }

\newcommand{\average}[2][0]{
    \setlength{\bracketsize}{0.75ex * #1 + 0.50ex}
    \ifthenelse
    { \equal{#1}{-1} }
    { \ensuremath{ {\left< #2 \right> } } }
    { \ensuremath{
        {               \left< \rule{0pt}{ \bracketsize} \right. \hspace{-1pt} }
        #2
        { \hspace{-1pt} \left. \rule{0pt}{ \bracketsize} \right>               }
} } }

\journal{Chinese Physics C}

\begin{document}

\begin{frontmatter}



\title{Holography and the internal structure of charmonium}


\author[a]{Nelson R. F. Braga}
\ead{braga@if.ufrj.br}

\author[b]{Yan F. Ferreira}
\ead{yancarloff@pos.if.ufrj.br}

\author[c]{William S. Cunha}
\ead{wscunha@pos.if.ufrj.br}

\affiliation[a,b,c]{
    organization={Instituto de Física, Universidade Federal do Rio de Janeiro}, 
    postcode={Caixa Postal 68528},
    state={Rio de Janeiro},
    country={Brazil}.
}

\begin{abstract}
Holographic models that consider classical vector fields in a 5-d background provide successful effective descriptions for heavy vector meson spectra. This holds both in the vacuum and in a thermal medium, like the quark gluon plasma. However, it is somehow mysterious the way that these phenomenological models work. In particular, what is the role of the fifth dimension and what is the relation between the holographic 5-d background and the physical (4-d) heavy mesons. Hadrons, in contrast to leptons, are composite particles with some internal structure, that depends on the energy at which they are observed. In this work, a static meson is represented by a heavy quark-antiquark pair with an interaction described by a Nambu Goto string living in the same 5-d background that provides field solutions leading to masses and decay constants of charmonium states. The interaction potential that shows up is linear for large distances with a string tension consistent with the effective Cornell potential. Introducing temperature $T$ in the background it is found, for the $J/\psi$ case, that there is a deconfining transition at some critical value of $T$. The results obtained indicate that the 5-d background is effectively representing the internal structure of the (static) charmonium (quasi) states.
\end{abstract}




\end{frontmatter}




\vspace{.75\baselineskip}
\section{Introduction}

Quarkonium states produced in heavy ion collisions are important sources of information about the quark gluon plasma (QGP). This state of strongly interacting matter, that behaves like a perfect fluid \cite{Bass:1998vz, Scherer:1999qq, Shuryak:2008eq, Casalderrey-Solana:2011dxg}, is not observed directly. The reconstruction of the QGP is based on the analysis of the final particles that reach the detectors in a heavy ion collision. One important tool is the analysis of suppression of charmonium states, relative to proton-proton collisions, that is interpreted as a consequence of dissociation in the medium \cite{Matsui:1986dk, Satz:2005hx}. This is the reason for the wide interest in understanding the behavior of quarkonium quasi-states in a thermal medium. 

The dissociation of quarkonium states in a plasma has been described in the recent literature by means of holographic models for heavy vector mesons inspired by gauge-string duality, like in Refs.\;\cite{Braga:2015jca,Braga:2016wkm, Braga:2017oqw,Braga:2017bml,Braga:2018zlu, Braga:2018hjt,Braga:2019yeh,MartinContreras:2021bis, Zollner:2021stb,Mamani:2022qnf}. 
In particular, the holographic model of Ref.\;\cite{Braga:2017bml} involves three energy parameters, associated with the heavy quark mass, the intensity of the strong interaction (string tension) and the scale of energy change in the  non-hadronic decay of quarkonium.  This model provides good estimates for the spectra of masses and decay constants and also for the dependence of the dissociation effect on temperature, magnetic field, density and angular momentum \cite{Braga:2017bml,Braga:2018zlu, Braga:2018hjt,Braga:2019yeh,Braga:2020myi,Braga:2021fey,Braga:2023fac}.

This phenomenological holographic model consists basically of a vector field living in a five-dimensional space with some background. Requiring the field to be normalizable, one finds a discrete set of solutions, corresponding to the meson states. It is important to remark that mesons are bound states of a quark-antiquark pair. The normalizable field solutions represent the different states. For charmonium, the solution with the smallest mass corresponds to $J/\psi$. The background of the model is engineered is such a way to provide the spectra of masses and decay constants. However, it lacks an interpretation for what is the relation between the five-dimensional phenomenological background and the interaction between the quark and the antiquark. The masses of the heavy meson states are a result of both the heavy quark masses and the quark-antiquark interaction. 

The question we will address in this work is: what is the relation between the 5-d background and the quark-antiquark interaction? Or, in other words, what is the relation between the background geometry and the internal structure of charmonium states? With this purpose, we study the holographic representation of the interaction between heavy quarks, consisting of a Nambu Goto string, in the model background. We start with the vacuum case and then analyze the finite temperature situation, when the metric develops a black hole (BH) geometry. The dissociation process of charmonium is then investigated from the point of view of the quark-antiquark interaction.

This work is organized in the following way. In section \ref{sec: model}, we review the holographic model for heavy vector mesons and present a different version, with a change in the sign of a quadratic term, that is necessary in order to have a confining background. In section \ref{sec: the string}, we study the quark-antiquark interaction by considering a string, with fixed endpoints, in the background of the model. Then, in section \ref{sec: finite temperature case}, the finite temperature case is considered and the dissociation processes is investigated. Some final remarks and conclusions are left for section \ref{sec: Conclusions}.


\vspace{.75\baselineskip}
\section{Holographic Model for Charmonium}
\label{sec: model}
 

    In gauge-string duality, the massless vector field \(V_m\) in the bulk is dual to the current operator \(j^\mu = \bar \psi \gamma^\mu\psi\), which represents a meson on the boundary (\(z \to 0\)). This correspondence provides a framework for describing mesons in terms of bulk fields. Based on this duality, a holographic model for quarkonium 
    was proposed in Ref.\;\cite{Braga:2017bml}, with an action integral of the form

\begin{gather}
    I = - \dfrac{1}{4 g_5^2} \int d^4x \int_{0}^{z_h}\, dz\, \sqrt{-{\bar g}}\, e^{-\phi(z)} {\bar g}^{\,mp} {\bar g}^{\,nq} F_{mn} F_{pq},
    \label{eq: original field action}
\end{gather}
 
with $F_{mn} =  \partial_m V_n - \partial_n V_m$.  
  The metric $\bar{g}$ is given by
\begin{gather}
    ds^2 = \bar{g}_{mn} dx^m dx^n = \dfrac{R^2}{z^2} \PAR[3]{\!-dt^2 + (dx^1)^2 + (dx^2)^2 + (dx^3)^2 + dz^2},
    \label{eq: metric 1}
\end{gather}
  where $x^1$, $x^2$ and $x^3$ are the spatial coordinates of the 4-d space where the meson lives and $z$ is the holographic coordinate, and the dilaton field is:
\begin{gather}
    \phi(z) = \kappa^2 z^2 + Mz + \tanh\!\PAR[3]{\dfrac{1}{Mz} - \dfrac{\kappa}{\sqrt{\Gamma}}}.
    \label{eq: dilaton tangent}
\end{gather}
The three parameters $\kappa, M $ and $ \Gamma $ are fixed in such a way to provide the best fit for the masses and decay constants of quarkonia states.

One can incorporate the dilaton $\phi$ in the metric through the transformation
\begin{gather}
   {\bar  g} = g\, e^{2\phi(z)}.
\end{gather}
The new metric $g$ is
\begin{gather}
    ds^2 = g_{mn} dx^m dx^n = \dfrac{R^2}{z^2} e^{-2\phi(z)} \PAR[3]{\!-dt^2 + (dx^1)^2 + (dx^2)^2 + (dx^3)^2 + dz^2},
    \label{eq: metric 2}
\end{gather}
and the action \eqref{eq: original field action} takes the form
\begin{gather}
    I = - \dfrac{1}{4 g_5^2} \int d^4x \int_{0}^{z_h}\, dz\, \sqrt{-{g}}\, {g}^{\,mp} {g}^{\,nq} F_{mn} F_{pq}.
    \label{eq: field action new metric}
\end{gather}

With the metric and the action written as in Eqs.\;\eqref{eq: metric 2} and \eqref{eq: field action new metric},
one can interpret the dilaton as being part of the geometry of the space. This interpretation does not affect the solutions for the vector fields, since actions \eqref{eq: original field action} and \eqref{eq: field action new metric} lead to the same equations of motion for the fields. However, in gauge-string duality, it is possible to represent the potential energy associated with the interaction between heavy quarks by a string connecting two fixed points on the boundary of the five-dimensional space and stretching into the bulk \cite{Maldacena:1998im, Kinar:1998vq}. The result depends on the metric of the space, so that metrics $g$ and $ \bar g $ lead to different potential energies. We will revise this point in section \ref{sec: the string}. If the potential energy increases linearly with the quark-antiquark distance, one has confinement. In order for this to happen,  the product of the metric components $g_{tt}$ and $g_{xx}$ must have a non-vanishing minimum, as we comment in section \ref{sec: the string} and as shown in detail in Ref.\,\cite{Kinar:1998vq}. This criteria is not satisfied by metrics \eqref{eq: metric 1} and \eqref{eq: metric 2}. However, if one changes the sign of the quadratic term in $z$ in metric \eqref{eq: metric 2} one finds confinement. For a similar discussion, in the context of the soft wall model, see \cite{Andreev:2006nw, Andreev:2006ct}.

If one uses the new dilaton
 \begin{gather}
    \phi(z) = -\kappa^2 z^2 - M z + \tanh\PAR{\frac{1}{M z} - \frac{\kappa}{\sqrt{\Gamma}}},
    \label{eq: new dilaton}
\end{gather}
instead of \eqref{eq: dilaton tangent} in the metric \eqref{eq: metric 2} when defining the metric $g$, it is still possible to fit the masses and decay constants of charmonium states and the corresponding quark-antiquark potential energy will increase linearly with the quark distance, showing confinement. From now on we consider the dilaton \eqref{eq: new dilaton}.

The action \eqref{eq: field action new metric} lead to the equations of motion
\begin{gather}
    \partial_n (\sqrt{-g}\, e^{-\phi} F^{mn}) = 0.
    \label{eq: eqs of motion in compact form}
\end{gather}

In order to describe a meson at rest, we choose a solution corresponding to zero spatial momentum  $V_m(t, \xx, z) = v_m(\omega,z) e^{-\complexi\omega t}$ and use the condition $V_z = 0$. Now, the equations of motion \eqref{eq: eqs of motion in compact form} become
\begin{align}
    \omega^2 v_j(\omega,z) - \PAR[3]{ \dfrac{1}{z} + \phi'(z) } v_j'(\omega,z) + v_j''(\omega,z) &= 0
    \qquad\qquad (j = 1,2,3),
    \label{eq: eq motion 123}\\
    - \PAR[3]{ \dfrac{1}{z} + \phi'(z) } v_t'(\omega,z) + v_t''(\omega,z) &= 0,
    \label{eq: eq motion t}\\
    v_t'(\omega,z) &= 0.
    \label{eq: eq motion z}
\end{align}
where the prime stands for the derivative with respect to $z$.

The equations \eqref{eq: eq motion t} and \eqref{eq: eq motion z} have the trivial solution $v_t = \mathrm{constant}$, and this constant must be zero in order to ensure normalization. The relevant equation is  \eqref{eq: eq motion 123}. Choosing a fixed polarization $\epsilon$, one can write $ v_i = \epsilon_i v $, with $\epsilon$ an unitary vector of the form  $(0,\epsilon_1, \epsilon_2, \epsilon_3, 0)$. Then, Eqs.\;\eqref{eq: eq motion 123} reduce to
\begin{align}
    \omega^2 v(\omega,z) - \PAR[3]{ \dfrac{1}{z} + \phi'(z) } v'(\omega,z) + v''(\omega,z) &= 0.
    \label{eq: eq motion}
\end{align}

Meson states are represented by normalizable solutions of the field. The normalization condition reads
\begin{gather}
    \int_{0}^{\infty} \dfrac{R}{z} e^{-\phi(z)}\, |v(\omega,z)|^2\, dz
    = 1,
    \label{eq: field normalization}
\end{gather}
that implies a boundary condition for the field:
\begin{gather}
    v(\omega,0) = 0.
    \label{eq: field boundary condition}
\end{gather}

The masses $m_n$ of the charmonium states are identified with the possible energy eigenvalues $\omega_n$ of the meson at rest, obtained by solving the equation of motion \eqref{eq: eq motion}  with the boundary condition \eqref{eq: field boundary condition}.

The decay constants are obtained from the equation \cite{Braga:2017bml}
\begin{gather}
    f_n = \frac{1}{g_5^2 m_n} e^{-\phi(0)} \lim_{z 
\rightarrow 0} \frac{R}{z} v'(\omega,z),
\end{gather}
where the constant $g_5$ is determined in the appendix \ref{ap: determination of the constant g5}.

The equation of motion \eqref{eq: eq motion}, with the dilaton field of \eqref{eq: new dilaton} has no analytical solution. If only the quadratic term is present  for which $\phi = \kappa^2 z^2$, corresponding to the soft wall model \cite{Karch:2006pv}, the solution is analytic and the  spectrum of masses is $m_n^2 = 4 \kappa^2 (n+1)$\; $(n = 0,1,2,3,\cdots)$, giving a nice description of light mesons. 

However, for heavy mesons, experimental data show that the mass of the state $n=0$ is considerably greater than the differences of mass $m_{n+1} - m_n$. This is a consequence of the fact that the masses of the heavy mesons have a very large contribution from the mass of the constituent quarks. In other words, the masses of  heavy mesons   depend not only on the strong interaction between the quarks but also on their masses. That is one of the reasons why one needs a dilaton with more parameters when dealing with heavy mesons.

Also, heavy meson decay constants should decrease with the number $n$. These feature is  not captured by the soft wall model \cite{Karch:2006pv}, where the decay constants do not depend on $n$.  In the tangent model, the solutions to the equation of motion \eqref{eq: eq motion} are numeric, but the fit of the masses and decay constants of heavy mesons is more precise.

With the definition $\psi(z) = \sqrt{R/z}\;e^{-\phi(z)/2}v(z)$, the equation of motion \eqref{eq: eq motion} is taken to the Schrödinger-like form
\begin{gather}
    -\psi'' + V\psi = \omega^2\psi,
\shortintertext{with}
    V = \frac{3}{4z^2} + \frac{1}{2 z}\phi' + \frac{1}{4}\phi'^2 - \frac{1}{2}\phi''
\end{gather}
This form of the equation can be used to explain the reason behind the dilatons \eqref{eq: dilaton tangent} or \eqref{eq: new dilaton} and why the fit of masses and decay constants is better with these dilatons. The hyperbolic tangent in \eqref{eq: dilaton tangent} or \eqref{eq: new dilaton} adds a potential well to the potential-like function $V$, which is responsible for rising the decay constants of the first states. The presence of a $\pm M z$ term on the dilaton adds a constant term to the potential, which guarantees a better fit for the masses of heavy mesons. The sign of this term was chosen in such a way that the potential in the Schrödinger-like equation leads to the best description of the charmonium states.

The best fit for the masses and decay constants is found using the following values for the model parameters
\begin{gather}
    \kappa = \SI{1.2}{\giga\electronvolt},  \qquad
    M = \SI{0.91}{\giga\electronvolt}  \qquad\text{and}\qquad
    \sqrt{\Gamma} = \SI{0.32}{\giga\electronvolt},
    \label{eq: parameters set 1}
\end{gather}
and the results obtained are shown on table \ref{tab: charmonium masses and decay constants}.  More details on this procedure of finding masses and decay constants can be found in Ref.\;\cite{Braga:2017bml}.

\begin{table}[htb]
\centering
\begin{tabular}{
    |m{1.0cm}<{\centering}
    |m{3.0cm}<{\centering}
    |m{2.8cm}<{\centering}
    |m{3.0cm}<{\centering}
    |m{2.8cm}<{\centering}
    |
}
    \hline
    \multicolumn{5}{|c|}{Charmonium Masses and Decay Constants} \\\hline
    State &
    Experimental Masses ($\si{\mega\electronvolt}$) &
    Masses on the tangent model ($\si{\mega\electronvolt}$) &
    Experimental Decay Constants ($\si{\mega\electronvolt}$) &
    Decay Constants on the tangent model ($\si{\mega\electronvolt}$)
    \\\hline
    $1S$  & $ 3096.900    \pm 0.006    $ & $2300$  & $ 416 \pm 4 $ & 411 \\\hline
    $2S$  & $ 3686.097    \pm 0.011    $ & $3445$  & $ 294.3    \pm 2.5    $ & 259 \\\hline
    $3S$  & $ 4040 \pm 4 $ & $4289$  & $ 187 \pm 8 $ & 206 \\\hline
    $4S$  & $ 4415 \pm 5 $ & $4982$  & $ 161 \pm    10 $ & 180 \\\hline
\end{tabular}\\[10pt]
\caption{Comparison of charmonium masses and decay constants obtained experimentally and from the tangent model.}
\label{tab: charmonium masses and decay constants}
\end{table}

The experimental values for the masses in Table \ref{tab: charmonium masses and decay constants} are taken from the Particle Data Group (PDG) \cite{ParticleDataGroup:2020ssz}. While the PDG does not directly  
show the decay constants $f_n$, it does provide the decay width $\Gamma_{n \to e^{+}e^{-}}$. The relationship between these quantities is given by the formula  \cite{Hwang:1997ie}
\begin{gather}
      f_n^2 \ = \dfrac{3 m_n}{4 \pi \alpha^2 c_V} \Gamma_{n \to e^{+}e^{-}},
\end{gather}
where $\alpha = 1/137$ is the fine structure constant; $c_V$ is the square of the charge of the quark, which, for charmonium, is $c_V = c_{J/\psi} = 4/9$; $m_n$ is the mass of the state whose radial excitation number is $n$; and $\Gamma_{n \to e^{+}e^{-}}$ is the decay width found on the PDG.

Let us define the root mean square percentage error (RMSPE) as
\begin{gather}
    \textrm{RMSPE}
    = 100\% \times \sqrt{ \dfrac{1}{N}
                     \sqrtsum{i\,=\,1}{N} \PAR[3]{\dfrac{y_i-\hat{y}_i}{\hat{y}_i}}^{\!\!2} },
\end{gather}
where $N = 8$ is the number of experimental points (4 masses and 4 decay constants),  the $y_i$'s are the values of masses and decay constants predicted by the model and the $\hat{y}_i$'s are the experimental values of masses and decay constants. With this definition, we have $\textrm{RMSPE} = 12.7\%$.


\vspace{.75\baselineskip}
\section{The String and the quark-antiquark potential}
\label{sec: the string}

In the context of the gauge-gravity duality, the interaction between two static color charges, or infinite mass quarks, is represented by a string connecting the quarks. The string stretches to the fifth dimension of the $\AdS_5$ space with a shape that minimizes its world sheet area or, equivalently, the corresponding Nambu-Goto action \cite{Maldacena:1998im, Kinar:1998vq}
\begin{align}
    \SNG &= \dfrac{1}{2\pi\alpha'} \int \sqrt{-g_{tt} g_{xx} dx^2 dt^2 - g_{tt} g_{zz}  dz^2 dt^2} \nonumber\\
         &= \dfrac{t}{2\pi\alpha'} \int dx \sqrt{-g_{tt} g_{xx} - g_{tt} g_{zz} (z')^2},
    \label{eq: nambu-goto action}
\end{align}
where $\alpha'$ is a constant with mass dimension $-2$. The string endpoints are located on the axis $x^1 \equiv x$ at the positions $ x = \pm r/2 $ and $r$ is the quark-antiquark distance, from the gauge theory point of view.

    This association was proposed by Maldacena in Ref.\;\cite{Maldacena:1998im} to show that the minimal area is proportional to the expectation value of the Wilson loop along a closed contour \(\mathcal{C}\), formed by the quark-antiquark pair at \(z \to 0\). This relationship can be expressed as follows: 
    \begin{gather}
        \langle W(\mathcal{C})\rangle \sim e^{-S_{NG}}.
    \end{gather}
    On the other hand, it is known from QCD that by taking the limit \(\Delta t\to\infty\), the Wilson loop of a quark-antiquark pair can be written as \(\langle W(\mathcal{C})\rangle \sim e^{-\Delta t E}\), thus giving \cite{Andreev:2006ct}
    \begin{gather}
        E = \frac{1}{\Delta t}S_{NG}.
    \end{gather}

\begin{figure}
    \centering
    \includegraphics[width=0.6\linewidth]{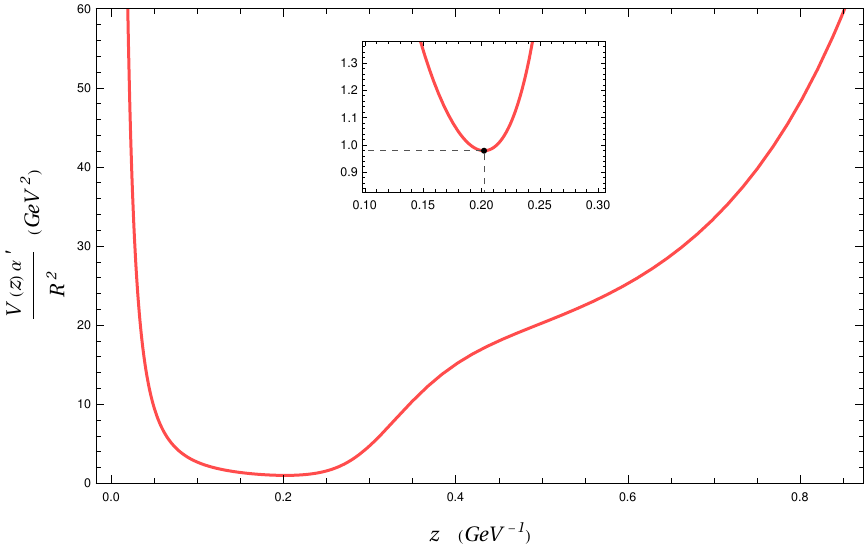}
    \caption{$V(z) \alpha'/ R^2$ as a function of the coordinate $z$}
    \label{fig: plot of V}
\end{figure}

Following Ref.\;\cite{Kinar:1998vq}, one can define
\begin{align}
    V(z) &=  \dfrac{1}{2\pi\alpha'} \sqrt{ - g_{tt} g_{xx} }= \dfrac{1}{2\pi\alpha'} \dfrac{R^2}{z^2} e^{-2\phi(z)}
    \label{eq: V at T=0}
\shortintertext{and}
    W(z) &=  \dfrac{1}{2\pi\alpha'} \sqrt{ - g_{tt} g_{zz}}  = \dfrac{1}{2\pi\alpha'} \dfrac{R^2}{z^2} e^{-2\phi(z)} = V(z).
    \label{eq: W at T=0}
\end{align}
The function $V(z)$ is plotted in Fig.\;\ref{fig: plot of V}. We will call $\zmin$ the point where the minimum of this function occurs. 

There is no time dependence on this problem. However, one can use a Hamiltonian type of formulation, but with the coordinate $x$  playing the role of the time, in order to minimize the functional $\SNG$. We define the Lagrangian
\begin{align}
    \calL(z,z',x) = \sqrt{V(z)^2 + W(z)^2 (z')^2},
    \label{eq: lagrangian}
\end{align}
which leads to the Hamiltonian\footnote{Observe that the Lagrangian and the Hamiltonian defined here have dimensions of $M^2$. They are not a Lagrangian or a Hamiltonian in standard definition used in classical mechanics.}
\begin{gather}
    \calH(z,p,x) = -\dfrac{V(z)^2}{\calL(z,z'(z,p),x)},
    \label{eq: hamiltonian}
\shortintertext{where}
    p = \dfrac{\delta\calL}{\delta z'} = \dfrac{W(z)^2 z'}{\calL(z,z',x)}
\end{gather}
is the conjugate momentum. Note that now we are using the prime to represent derivatives with respect to {\color{blue}the} coordinate $x$. 

As the Hamiltonian does not depend explicitly on $x$, its value is a constant of motion. Calling $z_0$ the point where the string crosses the $z$ axes, meaning that $z_0 = z(x=0)$, we write
\begin{align}
    \calH(z,p,x) = \calH(z,p) &= \at[3]{\calH(z,p(z,z'))}_{z\,=\,z_0,\, z'\,=\,0}\nonumber\\
                              &= -\dfrac{V(z_0)^2}{\sqrt{V(z_0)^2}} = - V(z_0) \equiv - V_0,
    \label{eq: hamiltonian 2}
\end{align}
since, for a symmetric (even) and smooth string we must have $z'(x=0) = 0$.

\begin{figure}[ht]
    \centering
    \begin{subfigure}[t]{.4\linewidth}
        \includegraphics[width=\linewidth]{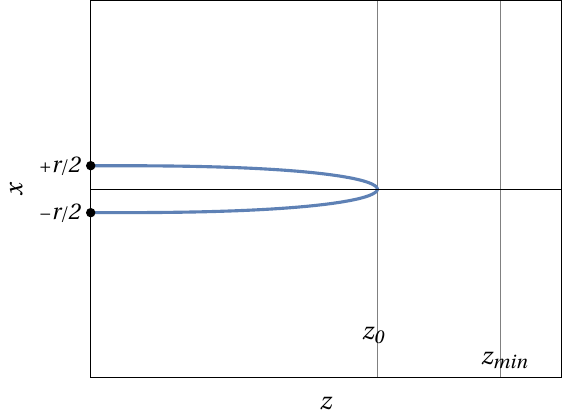}
        \caption{$ z_0 = 0.7\zmin $}
    \end{subfigure}\qquad
    \begin{subfigure}[t]{.4\linewidth}
        \includegraphics[width=\linewidth]{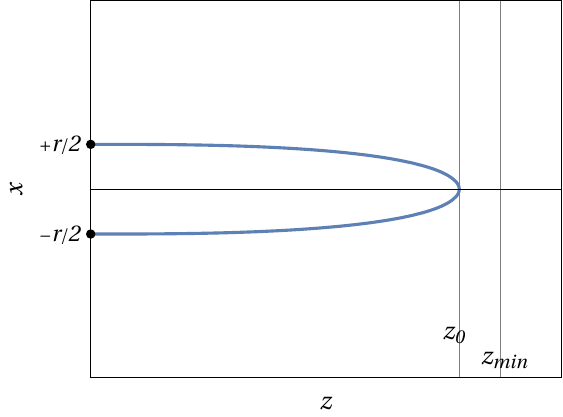}
        \caption{$ z_0 = 0.9\zmin $}
    \end{subfigure}\\[\baselineskip]
    \begin{subfigure}[t]{.4\linewidth}
        \includegraphics[width=\linewidth]{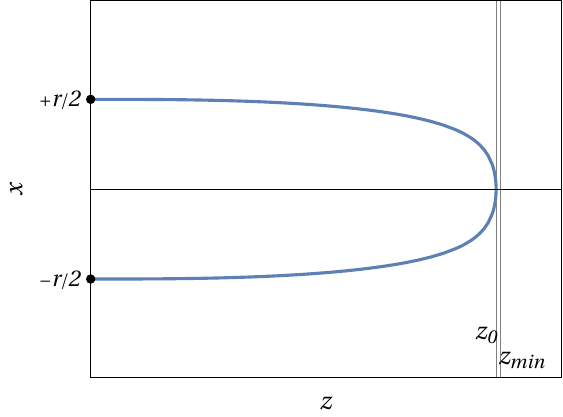}
        \caption{$ z_0 = 0.99\zmin $}
    \end{subfigure}\qquad
    \begin{subfigure}[t]{.4\linewidth}
        \includegraphics[width=\linewidth]{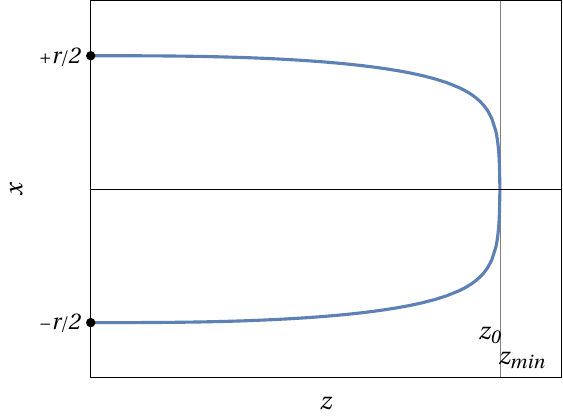}
        \caption{$ z_0 = 0.999\zmin $}
    \end{subfigure}
    \caption{Representation of four strings that are solutions of \eqref{eq: geodesic} with different values of $z_0$, corresponding to different distances $r$ between quarks.}
    \label{fig: geodesic}
\end{figure}

After some algebra, Eq.\;\eqref{eq: hamiltonian 2} leads to the equation of the geodesic
\begin{gather}
    z' = \pm \dfrac{V}{W} \sqrt{V^2/V_0^2 - 1},
    \label{eq: geodesic}
\end{gather}
for the $x < 0$ ($+$) and for the $x > 0$ ($-$) sides of the string. This defines the shape of the string. Some examples of this type of string, for different values of $z_0$, are shown in Fig.\;\ref{fig: geodesic}.

The string configuration is completely determined by the value of $z_0$ and the geodesic equation \eqref{eq: geodesic}. In particular, the distance $r$ between the quarks 
and the energy $ E$ of the string can be written in terms of $z_0$.

At this point one notes that $x$ and $z$ are real coordinates. So, $ z' = dz/dx $ cannot assume complex values. This implies that the argument of the square root on Eq.\;\eqref{eq: geodesic}, $ V^2/V_0^2 - 1 $, must be non-negative. This condition imposes a limit to the value of $z_0$. From Fig.\;1, one notices that the function $ V(z) $ is monotonically decreasing for $ z \le \zmin $. So, for strings that satisfy $ z_0 \le \zmin $, one has that $ V^2(z)/V_0^2 - 1 $ is always non negative. However, for $ z >  \zmin $, Fig.\;\ref{fig: plot of V} shows that $V(z) $ is monotonically increasing. So, for a string satisfying  $z_0 >  \zmin$, the factor $V^2(z)/V_0^2 - 1$ assumes negative values that would lead to a complex square root in Eq.\;\eqref{eq: geodesic}. Thus one concludes that the condition $ z_0 \le \zmin $ must hold. 


\subsection{Effective Potential}
\label{sec: Effective Potential}

The distance between the quarks is
\begin{gather}
    r = \int dx
      = 2 \int_{0}^{z_0} \dfrac{1}{z'}\, dz
      = 2 \int_{0}^{z_0} \dfrac{1}{\sqrt{V^2/V_0^2 - 1}} \dfrac{W}{V}\, dz.
    \label{eq: dist as function of z0}
\end{gather}
It is important to note that, as we will see in the sequence, when $ z_0 \to \zmin $ the quark separation diverges, $ r \to \infty $. 

The energy of the configuration is the length of the string
\begin{gather}
    E = \int \calL\, dx
      = 2 \int_{0}^{z_0} \dfrac{\calL}{z'}\, dz
      = 2 \int_{0}^{z_0} \dfrac{W}{\sqrt{1 - V_0^2/V^2}}\, dz,
    \label{eq: energy as function of z0}
\end{gather}
which is just the Nambu-Goto action \eqref{eq: nambu-goto action} divided by the time interval.

The expression \eqref{eq: energy as function of z0} is singular, since it includes the infinite masses of the quarks. One can regularize the energy following a similar approach as in Ref.\;\cite{Maldacena:1998im} of subtracting the contribution associated with two straight strings  
\begin{align}
    E &= 2 \int_{0}^{z_0^{}} \dfrac{W}{\sqrt{1 - V_0^2/V^2}}\, dz - 2 \int_{0}^{\zmin} W dz \nonumber\\[3pt]
      &= 2 \int_{0}^{z_0^{}} \PAR[4]{\dfrac{1}{\sqrt{1 - V_0^2/V^2}} - 1} W\, dz - 2 \int_{z_0^{}}^{\zmin} W dz.   
      \label{eq: regularized energy as function of z0}
\end{align}
The straight strings stretch from the boundary to the minimum of the metric factor $V(z)$, that occurs at $z = \zmin$.

Now, in order to compare the interaction energy of the quark-antiquark pair with the Cornell potential, we will analyze the asymptotic behaviors of the energy for large and small quark separation.  

\subsection{Potential Energy for large and small quark-antiquark separations}
\label{sec: Potential Energy for large and small quark-antiquark separations}

The asymptotic behavior of $r$ and $E$ for $z_0$ close to $\zmin$ was discussed for general metrics in \cite{Kinar:1998vq}. For the particular case of the present model, one has 
\begin{flalign}
    && r(z_0) &= 2\, \sqrt{ \dfrac{V(\zmin)}{V''(\zmin)} }\, \frac{W(\zmin)}{V(\zmin)} \ln\frac{1}{1-z_0/\zmin}\hspace{3em}
    && \llap{(\ensuremath{z_0 \approx \zmin}) \hspace{2em}}
    \label{eq: r for z0 close to zmin}
\shortintertext{and}
    && E(z_0) &= 2\, \sqrt{ \dfrac{V(\zmin)}{V''(\zmin)} }\, W(\zmin)\, \ln\frac{1}{1-z_0/\zmin}\hspace{3em}
    && \llap{(\ensuremath{z_0 \approx \zmin}). \hspace{1.7em}}
    \label{eq: E for z0 close to zmin}
\end{flalign}

Note, from Eq.\;\eqref{eq: r for z0 close to zmin}, that the region of $z_0$ close to $\zmin$ corresponds to the region of large $r$. Dividing \eqref{eq: E for z0 close to zmin} by \eqref{eq: r for z0 close to zmin}, we see that in this limit 
 \begin{flalign}
    && E (r) = V(\zmin)\, r
    && \llap{(large $r$). \hspace{1.8em}}
    \label{eq: E for large r}
\end{flalign}

This linear potential characterizes confinement. If it happens that, for a certain geometry, $V(\zmin) = 0$, the quarks of the dual gauge theory would be unconfined \cite{Kinar:1998vq}.

The Cornell potential \cite{Eichten:1974af}, that represents effectively the quark-antiquark interaction for the static case, has the form
\begin{gather}
    E(r) = -\frac{4}{3}\frac{\alpha_s}{r} + \sigma r.
    \label{eq: Cornell potential}
\end{gather}
With a similar linear term in $r$ in the large distance limit. Comparing Eqs.\;\eqref{eq: E for large r} and \eqref{eq: Cornell potential}, one finds that the string tension for the present holographic model is
\begin{gather}
    \sigma
    = V(\zmin)
    = \frac{R^2}{2\pi\alpha'}\frac{1}{\zmin^2} e^{-2\phi(\zmin)}.
    \label{eq: sigma from holography}
\end{gather}
On the other hand, for $z_0$ close to $0$, one finds \cite{Bruni:2018dqm}
\begin{flalign}
    && r(z_0) & = \frac{2\sqrt{\pi}}{3}\, \dfrac{\Gamma(7/4)}{\Gamma(5/4)}\, z_0
    && \llap{(\ensuremath{z_0 \approx 0}) \hspace{2em}}
    \label{eq: r for z0 close to 0}
\shortintertext{and}
    && E(z_0) & = \frac{2\sqrt{\pi}}{3} \,\dfrac{\Gamma(7/4)}{\Gamma(5/4)}\, \dfrac{R^2 e^{-2\phi(0)}}{\pi \alpha'}\, \frac{1}{z_0}
    && \llap{(\ensuremath{z_0 \approx 0}). \hspace{1.7em}}
    \label{eq: E for z0 close to 0}
\end{flalign}
From Eq.\;\eqref{eq: r for z0 close to 0}, we see that the region of $z_0$ close to $0$ corresponds to the region of small $r$. Therefore, multiplying \eqref{eq: E for z0 close to 0} by \eqref{eq: r for z0 close to 0}, we see that in this limit
\begin{flalign}
    &&
    E(r)
    = - \frac{(2\pi)^3}{\Gamma(1/4)^4}\, \frac{R^2 e^{-2\phi(0)}}{2\pi\alpha'}\, \frac{1}{r}
    && \llap{(small \ensuremath{r}). \hspace{1.8em}}
\end{flalign}
Again, this is in agreement, for the small $r$ limit, with the Cornell potential \eqref{eq: Cornell potential}. The predicted coupling constant $\alpha_s$ is
\begin{gather}
    \alpha_s = \frac{3}{4} \frac{(2\pi)^3}{\Gamma(1/4)^4} \frac{R^2 e^{-2\phi(0)}}{2\pi\alpha'}.
    \label{eq: alpha_s from holography}
\end{gather} 


\subsection{Numerical Results}
\label{sec: numerical results}

The parameters of the Cornell potential have been estimated considering a charmonium state as a static system of heavy quarks and solving the non-relativistic Schrödinger equation,
\begin{align}
    -\frac{1}{m_c}\nabla^2 \psi(\rr) + V(r) \psi(\rr) = E \psi(\rr),
\end{align}
for this two-body system (see, for example, \cite{Soni:2017wvy,Mateu:2018zym}). The eigenvalues $E_n$ of the energy correspond to the binding energies of the system. The masses of the various charmonium states have the form: $m_n = 2 m_c + E_n$, where $m_c$ is the mass of the charm quark. Using this approach, one is able to fix the parameters $\alpha_s$ and $\sigma$ of the Cornell potential \eqref{eq: Cornell potential} by fitting the experimental masses of charmonium states.

Following a similar approach, but using our holographic potential parametrized by Eqs.\;\eqref{eq: dist as function of z0} and \eqref{eq: energy as function of z0}, instead of the Cornell potential, we were able to fit the parameter $R^2/\alpha'$.
The results obtained were $R^2/\alpha' = 0.0226$  and the masses
shown on table \ref{tab: charmonium masses from the Shcrodinger equation}.

\begin{table}[htb]
\centering
\begin{tabular}{
    |m{1.0cm}<{\centering}
    |m{3.0cm}<{\centering}
    |m{3.0cm}<{\centering}
    |
}
    \hline
    \multicolumn{3}{|c|}{ Charmonium Masses from the Schrödinger equation}\\\hline
    State &
    Experimental Masses ($\si{\mega\electronvolt}$) &
    Masses obtained from the Shcrodinger equation ($\si{\mega\electronvolt}$)
    \\\hline
    $1S$  & $ 3096.900 \pm 0.006    $ & $3168$ \\\hline
    $2S$  & $ 3686.097 \pm 0.011    $ & $3652$ \\\hline
    $3S$  & $ 4040     \pm 4        $ & $4048$ \\\hline 
    $4S$  & $ 4415     \pm 5        $ & $4398$ \\\hline
\end{tabular}\\[10pt]
\caption{ Comparison of charmonium masses and decay constants obtained experimentally and from the Schrödinger equation.}
\label{tab: charmonium masses from the Shcrodinger equation}
\end{table}

 This approach for calculating the masses is not equivalent to the one used in section 2. In the ideal scenario of perfect equivalence, these masses would coincide with the ones shown on table \ref{tab: charmonium masses and decay constants}. One could consider fitting the parameter $R^2/\alpha'$ using the tangent model masses from table \ref{tab: charmonium masses and decay constants} instead of the experimental values. However, this method would propagate the errors inherent to the mass fitting within the tangent model.

Therefore, using Eqs.\;\eqref{eq: sigma from holography} and \eqref{eq: alpha_s from holography}, one obtains
\begin{equation}
    \sigma = \SI{0.163}{\giga\electronvolt^2}
    \qquad\qquad \text{and} \qquad\qquad
    \alpha_s = 0.00387.
\end{equation}
These are the estimates from the holographic model. The value of $\sigma$ is in a reasonable agreement with the ones obtained by the methods that apply the Cornell potential. For example, the authors of \cite{Ebert:2011jc, Soni:2017wvy, Mateu:2018zym} obtained $\sigma = \SI{0.18}{\giga\electronvolt^2}, \SI{0.18}{\giga\electronvolt^2}$ and $\SI{0.164}{\giga\electronvolt^2}$, respectively. The value of $\alpha_s$, on the other hand, is two orders of magnitude smaller than the typical values obtained from the Cornell potential. This result is consistent with the fact that the stringy description of the quark-antiquark interaction is expected to be appropriate for the large $r$ region, while $\alpha_s$ is related to the small $r$ region. 

\begin{figure}[ht]
    \centering
    \includegraphics[width=0.6\linewidth]{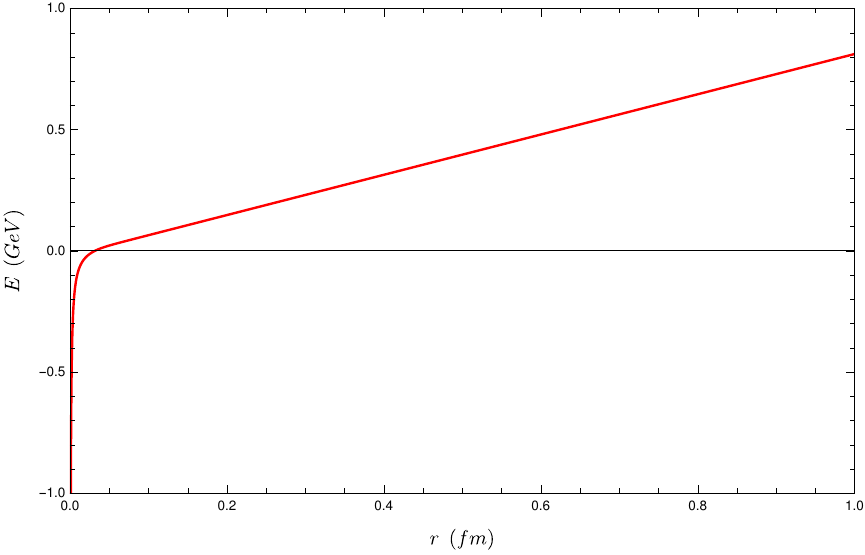}
    \caption{Energy of the string as a function of the distance between quark and antiquark.}
    \label{fig: E vs r at T = 0}
\end{figure}

Once the constant $R^2/\alpha'$ is fixed, we use Eqs.\;\eqref{eq: dist as function of z0} and \eqref{eq: energy as function of z0} to plot the curve $E(r)$ parametrized by $z_0$. This plot is shown in Fig.\;\ref{fig: E vs r at T = 0}. As expected from the asymptotic calculations, this graphic resembles the Cornell potential in the small and large $r$ limits. The linear growth behavior in the large $r$ region indicates confinement, as it implies that an infinite amount of energy would be required to separate the quarks.

\vspace{.75\baselineskip}
\section{Finite Temperature Case}
\label{sec: finite temperature case}

\subsection{Metric}
 
We introduce finite temperature by adding a black hole to our original $\AdS_5$ space \eqref{eq: metric 1}. The new metric is
\begin{gather}
    ds^2 = \bar{g}_{mn} dx^m dx^n = \dfrac{R^2}{z^2} \PAR[3]{\!-f(z) dt^2 + (dx^1)^2 + (dx^2)^2 + (dx^3)^2 + \frac{1}{f(z)}dz^2},
\shortintertext{with}
    f(z) = 1 - \frac{z^4}{z_h^4}.
\end{gather}
The fields that represent the mesons are subjected to the same action \eqref{eq: original field action}.
Then we introduce the dilaton into the metric, in  the same way as it was done in the zero temperature case, so that the new action assumes the same form of \eqref{eq: field action new metric} and the new metric becomes
\begin{gather}
    ds^2 = g_{mn} dx^m dx^n = \dfrac{R^2}{z^2} e^{-2\phi(z)} \PAR[3]{\!-f(z) dt^2 + (dx^1)^2 + (dx^2)^2 + (dx^3)^2 + \frac{1}{f(z)}dz^2}.
    \label{eq: metric with black hole}
\end{gather}
We assume that the dilaton parameters do not depend on the temperature.
The temperature of the plasma is identified with the Hawking temperature of this black hole, which is
\begin{align}
    T = \frac{1}{4\pi} \abs{f'(z_h)} = \frac{1}{\pi z_h}.
\end{align}
Inverting this equation, one can express the position of the horizon as a function of the temperature:
\begin{align}
    z_h = \frac{1}{\pi T}.
\end{align}


\subsection{Dissociation}

Using the metric \eqref{eq: metric with black hole}, we generalize the definition of the function $V(z)$ to include the dependence on the temperature:
\begin{equation}
    V(T, z) =  \dfrac{1}{2\pi\alpha'} \sqrt{ - g_{tt} g_{xx}} = \dfrac{1}{2\pi\alpha'} \dfrac{R^2}{z^2} e^{-2\phi(z)}\sqrt{f(z)}.
    \label{eq: V at finite T}
\end{equation}
Note that the function $f(z)$ depends on $z_h$ and consequently on $T$. On the other hand, the function
\begin{align}
    W(z) =  \dfrac{1}{2\pi\alpha'} \sqrt{ - g_{tt} g_{zz} } = \dfrac{1}{2\pi\alpha'} \dfrac{R^2}{z^2} e^{-2\phi(z)},
    \label{eq: W at finite T}
\end{align}
does not change, since the factors of $f(z)$ in $g_{tt}$ and $g_{zz}$ cancel out.
    
\begin{figure}[ht]
    \begin{subfigure}[h]{0.48\textwidth}
        \includegraphics[width=1\linewidth]{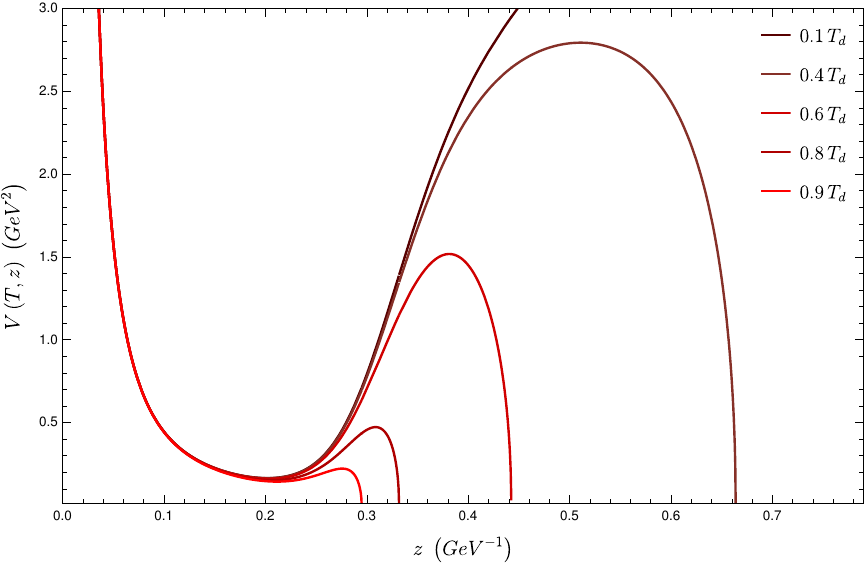}
         \end{subfigure}
    \hfill
    \begin{subfigure}[h]{0.48\textwidth}
        \includegraphics[width=1\linewidth]{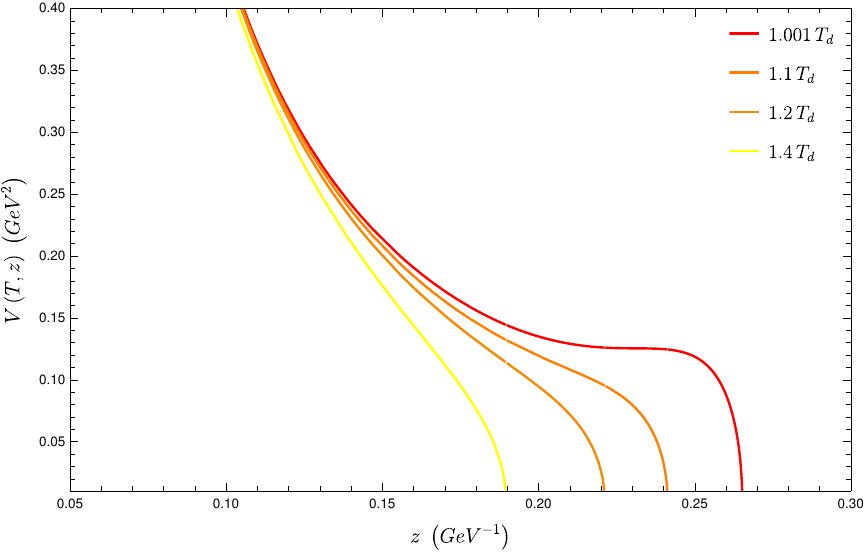}
                   \end{subfigure}
    \caption{ Some illustrative plots of $V(z)$ for different temperatures. On the left panel, $ T < T_d$.  On the right panel, $ T > T_d$. }
    \label{fig: V2 as function of temperature}
\end{figure}

The variation of the function $V(T, z)$ with the coordinate $z$ is strongly affected by the temperature. For low temperatures, $V$ presents two local minima: one, that is similar to the case of zero temperature, is a non-vanishing minimum at some point $z = \zmin(T)$; the other one is the zero of the function $V$ at $z = z_h(T)$. As a consequence,  $V$ also has a local maximum, located  between $\zmin$ and $z_h$. As the temperature increases, this picture  changes. For temperatures greater than some value $T = T_d$, $V$ has only one minimum, which is zero, at $z = z_h(T)$. In this case, the local non-vanishing minimum disappears, as well as the local maximum.

In Fig.\;\ref{fig: V2 as function of temperature}, plots of $V(T, z)$ are shown for different values of the temperature. On the left panel, one finds temperatures $ T < T_d$, while the right panel presents cases with $ T \geq T_d $. These plots illustrate the different behaviors. When the temperature grows, starting from a small value $ T < T_d $, the height of the local maximum decreases and its position approaches the point $\zmin(T)$. The temperature $ T_d $ is the one where both of these points, minimum and maximum, coincide, becoming an inflection point.

As we discussed on Sec.\;\ref{sec: Potential Energy for large and small quark-antiquark separations}, the minimum on $z$ of the function $V(T,z)$ is the key to determine whether there is confinement or not. For a fixed $T \ge T_d$, the function $V$ is monotonically decreasing with respect to $z$, with only one local minimum at $z = z_h$, where $V(z_h) = 0$. This indicates the absence of confinement. For that reason, we call $T_d$ the dissociation temperature. It can be determined graphically by the disappearance of the   non-vanishing   minimum on plots like those on Fig. \ref{fig: V2 as function of temperature}, or by imposing that the local non-vanishing minimum and local maximum to occur at the same point, that, as explained before, becomes an inflection point.   

If one uses the set of parameters shown in \eqref{eq: parameters set 1}, the dissociation temperature obtained from this method would be $T_d = \SI{1.2}{\giga\electronvolt}$. This value is unsatisfactorily larger than the predictions of lattice theory \cite{Karsch:2005nk} that indicate dissociation of $J/\psi $ at about $ 1.5 T_c$,  where $T_c$ is the critical temperature of the QGP formation by the dissociation of the light flavor hadrons. This corresponds to a dissociation temperature of the order of $ \, \sim \SI{0.25}{\giga\electronvolt}$.

This result motivates a different procedure in order to fix the the model parameters, that is to include the dissociation temperature as one of the physical quantities to be fitted, besides the masses and the decay constants. Following this approach, one finds the new set of  parameters:
\begin{gather}
    \kappa = \SI{1.1}{\giga\electronvolt},  \qquad
    M =  
    \SI{0.11}{\giga\electronvolt}\qquad\text{and}\qquad
    \sqrt{\Gamma} = \SI{0.26}{\giga\electronvolt}.
    \label{eq: parameters set 2}
\end{gather}
These parameters produce the results of masses and decay constants shown in table \ref{tab: charmonium masses and decay constants 2} and result in a reasonable value for the dissociation temperature
\begin{equation}
T_d = \SI{316}{\mega\electronvolt}.
\end{equation}
The errors in the decay constants increase considerably with this new approach. The new root mean square percentage error is $\mathrm{RMSPE} = 23\%$, with the temperature included in the calculation.
\begin{table}[htb]
\centering
\begin{tabular}{
    |m{1.0cm}<{\centering}
    |m{3.0cm}<{\centering}
    |m{2.8cm}<{\centering}
    |m{3.0cm}<{\centering}
    |m{2.8cm}<{\centering}
    |
}
    \hline
    \multicolumn{5}{|c|}{Charmonium Masses and Decay Constants} \\\hline
    State &
    Experimental Masses ($\si{\mega\electronvolt}$) &
    Masses on the tangent model ($\si{\mega\electronvolt}$) &
    Experimental Decay Constants ($\si{\mega\electronvolt}$) &
    Decay Constants on the tangent model ($\si{\mega\electronvolt}$)
    \\\hline
    $1S$  & $ 3096.900    \pm 0.006    $ & $2399$  & $ 416 \pm 4 $ & 298 \\\hline
    $2S$  & $ 3686.097    \pm 0.011    $ & $3560$  & $ 294.3    \pm 2.5    $ & 258 \\\hline
    $3S$  & $ 4040 \pm 4 $ & $4011$  & $ 187 \pm 8 $ & 239 \\\hline
    $4S$  & $ 4415 \pm 5 $ & $4590$  & $ 161 \pm    10 $ & 229 \\\hline
\end{tabular}\\[10pt]
\caption{Comparison of charmonium masses and decay constants obtained experimentally and from the tangent model with parameter that give the best fit of the dissociation temperature.}
\label{tab: charmonium masses and decay constants 2}
\end{table}

Let us now analyze the behavior of the string and of the free energy in some detail. The distance between the quarks has the same form as in the zero temperature case:
\begin{gather}
    r(T, z_0)
    = \int dx
    = 2 \int_{0}^{z_0} \dfrac{1}{z'}\, dz
    = 2 \int_{0}^{z_0} \dfrac{1}{\sqrt{V^2/V_0^2 - 1}} \dfrac{W}{V}\, dz,
    \label{eq: dist as function of z0 at finite T}
\end{gather}
but now $V$ is a function of both $T$ and $z$. Since the string behavior is different, depending on the temperature been lower or higher than $T_c$,  let us analyze these situations separately. 

\subsection{Low temperatures: $ T < T_d$ }

At finite temperature,   one can consider an extension of the proposal of Maldacena, discussed in section \ref{sec: the string}. Now, the Nambu-Goto action is related to Polyakov loops  
        \begin{gather}
            \langle\mathcal{P}(\vec x_1)\mathcal{P}(\vec x_2)\rangle \sim e^{-S_{NG}(T)}.
        \end{gather}
         As in the zero-temperature case, this expectation value is known and, in this scenario, is proportional to the free energy\footnote{We have ignored normalization terms.}
        \begin{gather}
            \langle\mathcal{P}(x_1)\mathcal{P}(x_2)\rangle \sim e^{-\frac{1}{T}F}.
        \end{gather}
        This leads to
        \begin{gather}
            F = T S_{NG}.
        \end{gather}
        
        Note that the string that minimizes the area of the world sheet also minimizes the free energy of the system. Then using \(\Delta t = 1/T\), we write
        
\begin{align}
    F(T, z_0)
    &= 2 \int_{0}^{z_0^{}} \dfrac{W}{\sqrt{1 - V_0^2/V^2}}\, dz - 2 \int_{0}^{\zmin(T=0)} W dz \nonumber\\[3pt]
    &= 2 \int_{0}^{z_0^{}} \PAR[4]{\dfrac{1}{\sqrt{1 - V_0^2/V^2}} - 1} W\, dz - 2 \int_{z_0^{}}^{\zmin(T=0)} W dz,
    \label{eq: regularized free energy as function of z0}
\end{align}
where the same regularization of the zero temperature case was used.

    \begin{figure}[ht]
        \begin{subfigure}[h]{0.48\textwidth}
            \includegraphics[width=1\linewidth]{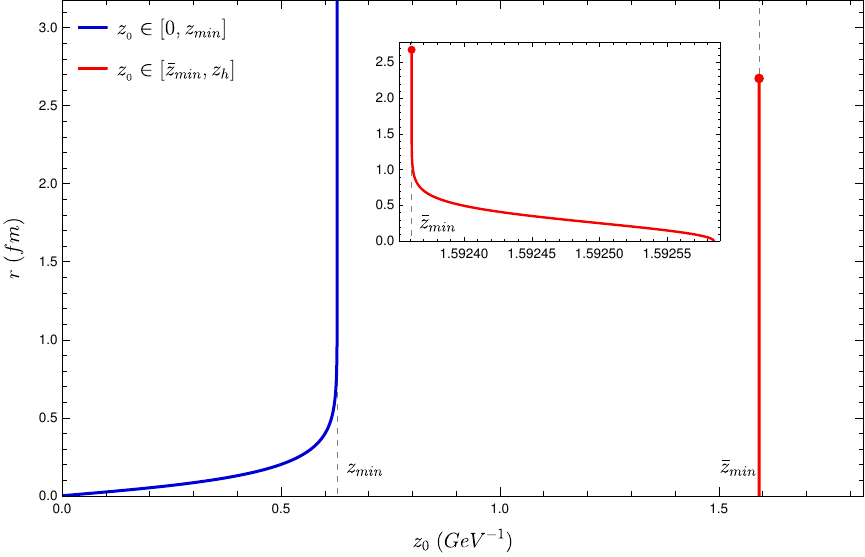}
        \end{subfigure}
        \hfill
        \begin{subfigure}[h]{0.48\textwidth}
            \includegraphics[width=1\linewidth]{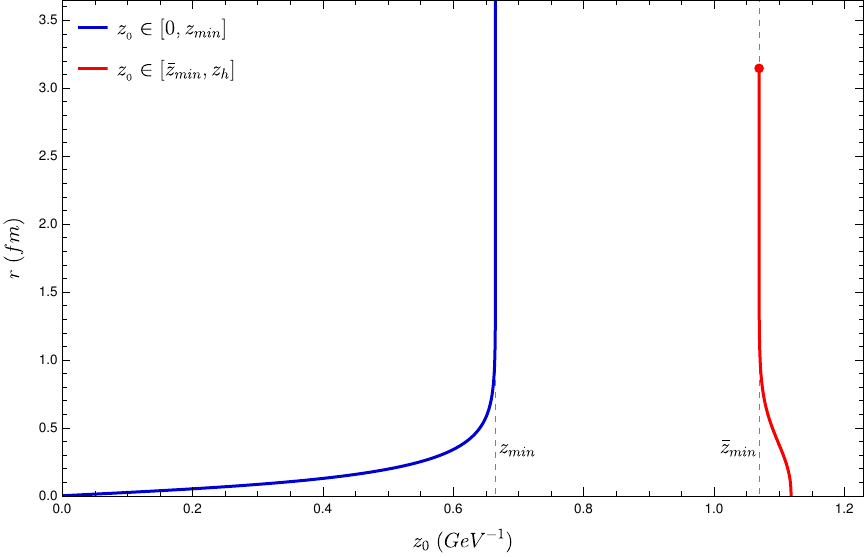}
        \end{subfigure}
        \caption{ Quark antiquark distance as a function of $z_0$.  Left panel: \(T=0.6T_d\). Right panel: \(T=0.95T_d\).}
        \label{fig: r as function z0}
    \end{figure}

At zero temperature, we saw that the domain of possible values of $z_0$ is $(0, \zmin)$. This is necessary in order to prevent the quantity $V^2/V_0^2 - 1$ of being negative, since it is inside a square root in the geodesic equation \eqref{eq: geodesic}. At finite temperature, for $ 0 < T < T_d $, $z_0$ can be in two different regions:\footnote{To keep the notation clean, we are omitting the dependence on $T$ of $\zmin(T)$, $\zminbar(T)$ and $z_h(T)$.}
$(0, \zmin)$ and $( \zminbar, z_h)$, where $\zminbar$ is the point satisfying  $ \zminbar > \zmin$ such that $V(\zminbar) = V(\zmin)$. One can understand the situation by looking at the left panel of Fig. \ref{fig: V2 as function of temperature}. 

So, in principle, one should have to consider more than one string configuration for each temperature. However, we will see that one of them is dominant. Applying Eq.\;\eqref{eq: dist as function of z0 at finite T} to the first region, $(0, \zmin)$, we find that the distance between quarks, \(r\), increases continuously from zero, going to infinity when \(z_0 \to \zmin\). This behavior is similar to the one found at the zero temperature. On the other hand, in the second region, \(r\) starts from a finite value at \(z_0 = \zminbar(T)\) and goes to zero at \(z_0 = z_h(T)\). These two behaviors can be seen in Fig. \ref{fig: r as function z0}.

\begin{figure}[ht]
    \centering
    \includegraphics[width=0.6\linewidth]{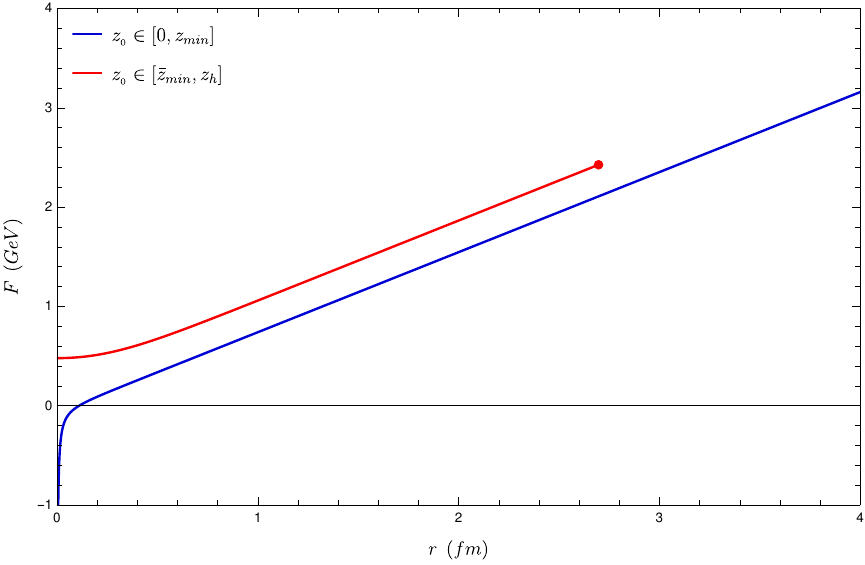}
    \caption{Free energy $F$ as function of the distance between the quarks $r$. The upper curve (red) corresponds to strings with $z_0$ in the region $(0, \zmin(T))$ and the lower one (blue), to strings with $z_0 \in (\zminbar(T), z_h(T)]$. In both cases, \(T = 0.8 T_d\).}
    \label{fig: the two energy regions}
\end{figure}

\begin{figure}[ht]
    \centering
    \includegraphics[width=0.75\linewidth]{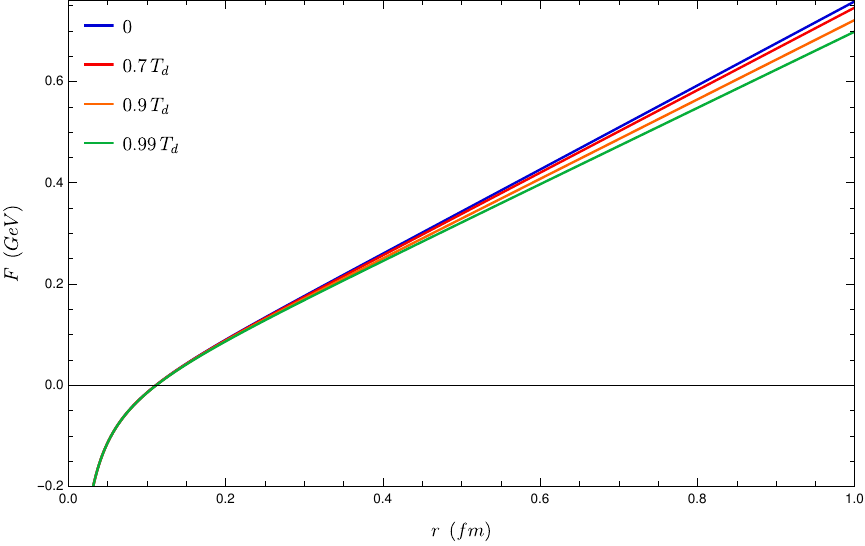}
    \caption{Free energy $F$ as a function of the quark separation $r$ for four illustrative temperatures: $ T = 0; T=  0.7\, T_d; T =  0.9\, T_d$ and $ T =  0.99\, T_d $.}
    \label{fig: various temperatures below Td}
\end{figure}

Now, using Eq.\;\eqref{eq: regularized free energy as function of z0} for each of the two regions for a given temperature \(T\) and making the parametrization \(\left(r(T, z_0), F(T, z_0)\right)\), we find the free energy profiles plotted in Fig. \ref{fig: the two energy regions}. Note that for an $r \in \left(0, r(\zminbar)\right]$ there are two possibilities of energy for each $r$, one corresponds to a string with $z_0 \in (0, \zmin)$ and the other corresponds to a string with $ z_0 \in [\zminbar, z_h] $. However, the string assumes the configuration that minimizes the Nambu-Goto action. Since \(F(T, z_0)\) is proportional to the action, the dominant configuration is the one that minimizes \(F(T, z_0)\) for a given \(r\).
For this reason, only the strings with $z_0$ in the region $(0, \zmin)$ will be considered. The result for the dependence of the free energy on the quark-antiquark separation $r$ is that of a confining potential again. The form is similar to the one found in the zero temperature case, with a linear behavior for large $r$.
A small difference that is worth pointing is that the string tension slightly reduces when temperature grows. The result is illustrated in Fig.\;\ref{fig: various temperatures below Td}.

\subsection{High temperatures: $T > T_d$}
\begin{figure}[t]
    \begin{subfigure}[h]{0.48\textwidth}
        \includegraphics[width=1\linewidth]{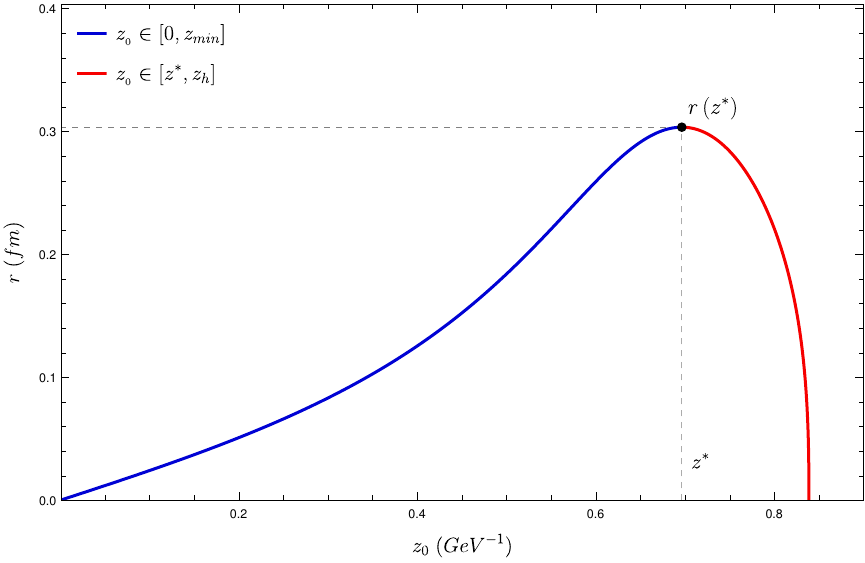}
    \end{subfigure}
    \hfill
    \begin{subfigure}[h]{0.48\textwidth}
        \includegraphics[width=1\linewidth]{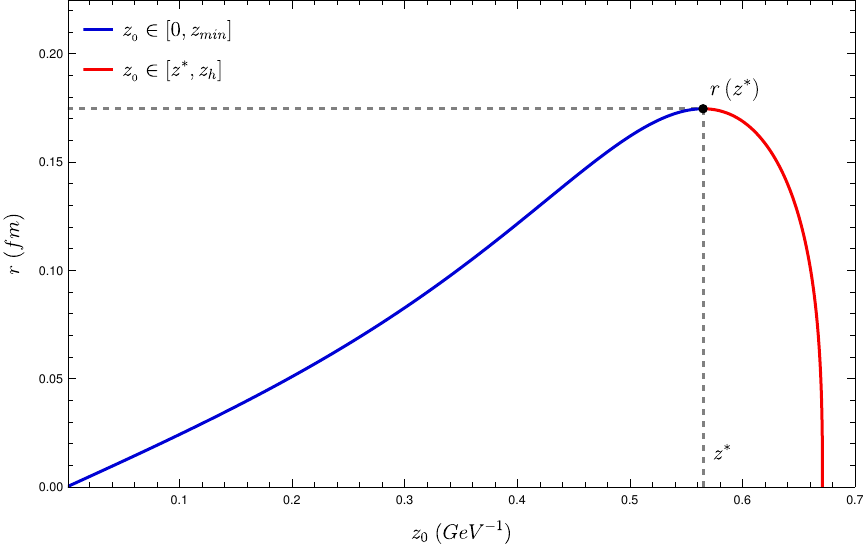}
    \end{subfigure}
    \caption{ Quark distance as a function of the maximum value $z_0$ of coordinate $z$. Regions: $z_0 < z^*$ (blue) and $z_0 > z^*$ (red). Temperatures \(T=1.2T_d\) (left panel) and \(T=1.5T_d\) (right panel).} 
    \label{fig: distance as function of z0 above Td}
\end{figure}

For $ T > T_d $, $\zmin$ does not exist anymore, so that the function \(V(T, z)\) has only one minimum at \(z=z_h(T)\).
In this situation, the distance between quarks, \(r(z_0)\), does not go to infinity, but has a maximum value at some \(z_0 = z^*\), from which $r(z_0)$ decreases until it reaches the value zero at \(z_0 = z_h(T)\). One can seen this in Fig. \ref{fig: distance as function of z0 above Td}.

\begin{figure}[ht]
    \centering
    \includegraphics[width=0.6\linewidth]{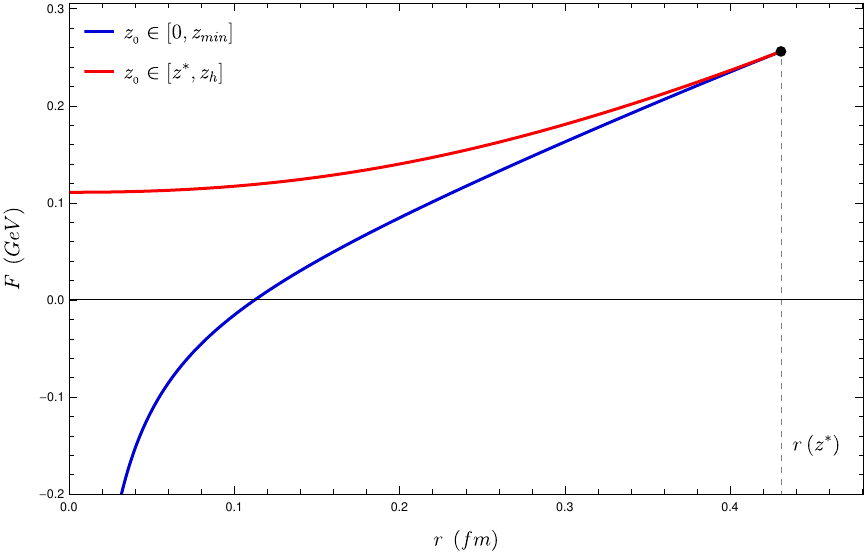}
    \caption{ Free energy as a function of the quark distance at  \(T = 1.01T_d\)}
    \label{fig: two region of energy above Td}
\end{figure}

The energy \(F(T, z_0)\) also presents a maximum finite value at $z_0 = z^*$. For higher values of $z_0$, $F$ decreases continuously until it vanishes at \(z_0 = z_h(T)\). Using  the parametrization \((r(T, z_0), F(T, z_0))\) for some temperature \(T > T_d\), we plot the free energy as a function of the distance between the quarks in Fig. \ref{fig: two region of energy above Td}.

As one can see from Fig. \ref{fig: distance as function of z0 above Td}, the region \(z_0 \in \left[ z^*, z_h \right]\) at $T > T_d$ plays a similar role as the region \(z_0 \in \left[ \zminbar, z_h\right]\) did in the case of temperatures below \(T_d\). In both cases, $r$ starts from a finite value and decreases to zero. Also, from Fig. \ref{fig: two region of energy above Td}, we see that when comparing two strings with the same distance between quarks, the one with $z_0 \in [z^*, z_h]$ has a value of free energy that is larger than the one for the string with $z_0 \in [z^*, z_h]$. As the string must minimize the free energy, the configuration with $z_0 \in [z^*, z_h]$ is not formed, so that we can disregard this region, similarly to what happened with the region of $z_0 \in [\zmin, z_h]$ in the case with temperature below $T_d$.

\begin{figure}[ht]
    \centering
    \includegraphics[width=0.4\linewidth]{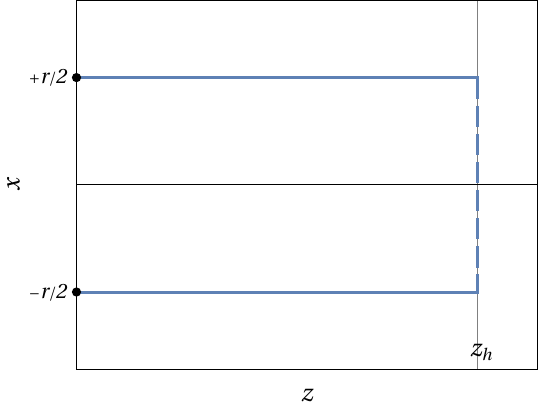}
    \caption{  
    String configurations corresponding to free quarks. This configuration is only formed for $r > r_d$, and, above $T>T_d$, the energy associated to this configuration do not depend on $r$.
    }
    \label{fig: deconfined representation}
\end{figure}

Therefore, we obtain a Cornell like potential again, but this time with a maximum range of \(r(T, z^*)\). From this point on, the quarks become free. To represent this situation, we have to consider a third configuration: two lines, which represent the two quarks, going straight in the $z$ direction up to \(z_h\), as sketched in Fig.\;\ref{fig: deconfined representation}. Note that these lines do not involve any variation along the $x$ coordinate. Therefore, the corresponding energy is independent of the value of the quark separation $r$. This is what characterizes freedom in a two body problem: changing the distance between these bodies does not require any energy cost.

This configuration requires a different treatment to obtain the energy. Let us consider the Nambu-Goto action but now for strings that are straight lines in the $z$ direction
\begin{align}
    \SNG &= \frac{t}{2\pi\alpha'}\int^{z_h}_0 dz\sqrt{-g_{tt}g_{zz}} \nonumber\\
    &= t \int^{z_h}_0 W(z)dz,
\end{align}
where we used the definition \eqref{eq: W at finite T}. The time interval $t$ has a trivial role for this static string. 

Then, the energy of the set of two straight lines, which represents two free quarks, is given by
\begin{equation}\label{eq: free constant energy}
    F_{\infty }(T) = 2 \int_{\epsilon(T)}^{z_h(T)} W(z) dz,
\end{equation}
where, in order to avoid the singularity at $z = 0$, we introduced a regularization parameter \(\epsilon(T)\) that depends on the temperature. We numerically set it to a value such that the energy of the non-interacting quark pair coincides with the maximum energy \(F (r(z^*), T)\), in a similar way as it was done in \cite{Colangelo:2010pe}. In other words, the parameter \(\epsilon(T)\) is chosen in such a way the free energy is a continuous function of $r$.

\begin{figure}[ht]
    \centering
    \includegraphics[width=0.75\linewidth]{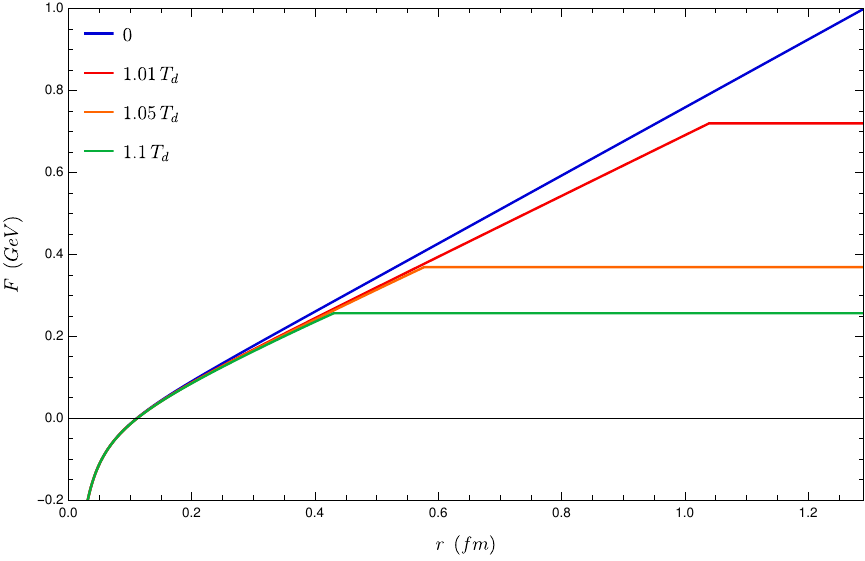}
    \caption{Interaction energy for three temperatures larger than $T_d$. The energy for the case $T=0$ is also shown, for comparison.}
    \label{fig: full potential at finite temperature}
\end{figure}

The physical picture that emerges from this approach is non-trivial and very interesting. For temperatures larger than \(T_d\), we have an interaction given by a potential similar to Cornell's potential for small values of $r$. However, when the quark distance is larger than the value \(r(T, z^*)\), the energy ceases to vary with $r$. The corresponding string assumes the configuration of two straight lines going to \(z_h\), with energy given by Eq.\;\eqref{eq: free constant energy}.  The string configuration corresponding to the unconfined quarks is depicted in Fig.\;\ref{fig: deconfined representation}. The energy associated to this configuration $F_\infty$ does not depend on the distance $r$. The resulting potential is plotted in Fig.\;\ref{fig: full potential at finite temperature} for some illustrative temperatures.

The potential obtained is similar to the Schwinger potential predicted by QCD for quark deconfinement in the presence of temperature \cite{sarkar2009physics, satz2012extreme, Allton:2013uua}. The main difference is that in the Schwinger potential, there is a smooth transition, characterizing a crossover, whereas here we observe a discontinuous transition, characterizing a first-order transition.

\begin{figure}[ht]
    \centering
    \includegraphics[width=0.6\linewidth]{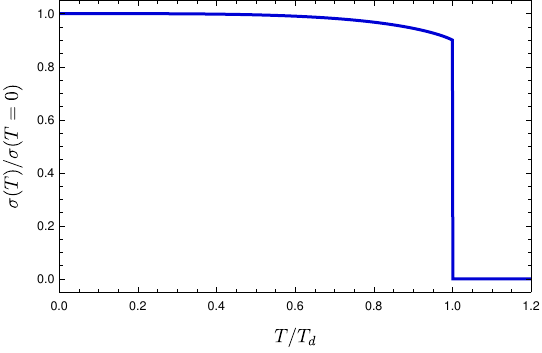}
    \caption{String tension variation with the temperature.  }
    \label{fig: V with temperature}
\end{figure}

If we analyze large distances, it is clear that for any temperature above \(T_d\), the mesons will be deconfined, since in this limit the string tension $\sigma(T) = V(T, \zmin) = 0 $. The variation of $ \sigma(T)$ is shown in Fig.\;\ref{fig: V with temperature}.    Note that for this system, \(\sigma(T)\) serves as an order parameter for the confined-deconfined phase transition, being identically zero in the deconfined phase and non-zero in the confined phase. However, in contrast to lattice QCD, which indicates a continuous decrease of \(\sigma(T)\) to zero at \(T_d\) \cite{Cardoso:2011hh, Bicudo:2011hn}, here we observe a discontinuous jump to zero at \(T_d\), indicating the first-order transition.

\begin{figure}[ht]
    \begin{subfigure}[h]{0.48\textwidth}
        \includegraphics[width=1\linewidth]{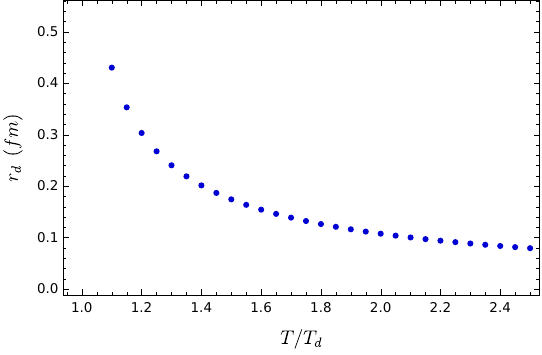}
        \caption{}
        \label{fig: distance max}
    \end{subfigure}
    \hfill
    \begin{subfigure}[h]{0.48\textwidth}
        \includegraphics[width=1\linewidth]{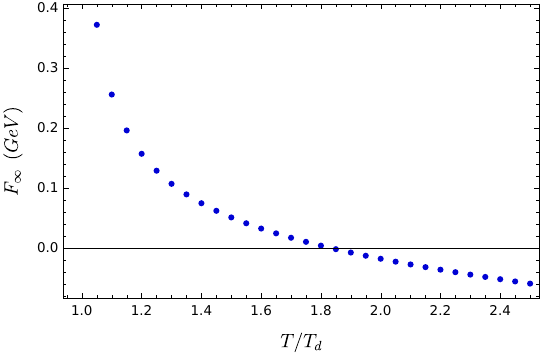}
        \caption{}
        \label{fig: energy max}
    \end{subfigure}
    
     \caption{                  
        (a) The maximum range between the interaction between quarks as a function of temperature. (b) Free energy of the deconfined phase as a function of temperature.}
    \label{fig: energy and distance max}
\end{figure}

    Looking more closely at Fig.\;\ref{fig: full potential at finite temperature}, one can observe that as the temperature increases, the maximum range of interaction decreases due to the screening effect of the medium. This maximum range is known in the literature as the Debye radius \(r_d\). In Fig.\;\ref{fig: distance max}, we show its dependence on temperature. Another important observation from Fig.\;\ref{fig: full potential at finite temperature} is the decrease in the free energy of the deconfined phase \(F_\infty(T)\) with temperature, as shown in more detail in Fig.\;\ref{fig: energy max}. These results are qualitatively in agreement with results from lattice QCD in \cite{digal2005heavy, Bicudo:2011hn, Allton:2013uua, Cardoso:2011hh}. 
    
    Also note that in figure \ref{fig: energy max}, a region of negative energy appears for \textit{very high} temperatures, corresponding to complete dissociation, where the free quarks exhibit a Coulomb-like interaction \cite{satz2012extreme}.

\section{Conclusions}
\label{sec: Conclusions}

The zero temperature  spectra of masses and decay constants as well as the thermal properties of quarkonium quasistates in a plasma have been successfully described recently by means of holographic models like \cite{Braga:2018zlu, Braga:2018hjt,Braga:2019yeh,MartinContreras:2021bis, Zollner:2021stb,Mamani:2022qnf}. In these models the heavy mesons are represented by a vector field living in a five-dimensional space with some background. The main objective of the present work was to shed some light into the reasons why this kind of model is able to reproduce quarkonium properties. 

In particular, we aimed to find an interpretation for the role of the extra dimension, with some non flat geometry, in the holographic models of heavy vector mesons. The guiding line of our study was the fact that mesons are not elementary particles in the strict sense,  but have an internal structure. A static meson can be viewed as consisting of a strongly interacting quark-antiquark pair. From the holographic point of view, the interaction of static color sources can be represented by a string living in a five-dimensional background  with endpoints fixed at the position of the sources. The interaction energy of the quarks is proportional to the world sheet area. For the charmonium case, we proposed to adapt the procedure of Refs.\;\cite{Maldacena:1998im, Kinar:1998vq} to a background geometry corresponding to the one used in holographic models that successfully describe masses,  decay constants and the thermal behavior of quarkonium states (or quasi-states) \cite{Braga:2017bml, Braga:2018zlu, Braga:2018hjt,Braga:2019yeh}.

It was necessary to make a change in the model of \cite{Braga:2017bml}, because the associated background would not be confining and therefore not appropriate to describe hadrons. In order for a geometry to represent confinement, it is necessary that the product of the metric components in the time and in the transverse directions have a non-vanishing minimum. 
Using a set of parameters that fit masses and decay constants and also provides a confining potential for the quark- antiquark interaction, we found a potential that has the asymptotic form of the well known Cornell potential. For the limit of large quark separation, we obtained a result for the string tension compatible with the values considered in the Cornell potential in the literature. 

Considering the finite temperature case, it was found that the string tension decreases with the temperature up to some dissociation temperature $T_d$. 
For higher temperatures $ T > T_d $, the linear term in the potential disappears for distances greater than some value that decreases with the temperature, as it is shown in Fig. \ref{fig: full potential at finite temperature}. The dissociation temperature initially obtained was too high compared to lattice predictions. However, it was shown that it is possible to obtain a new set of parameters that provide a $T_d$ compatible with expectations. 

I summary, we have show, in a qualitative way, that one can associate the five dimensional background of a previous successful holographic model for charmonium as representing the internal structure of this composite particle.


\appendix

\vspace{.75\baselineskip}
\section{Determination of the constant $g_5$}
\label{ap: determination of the constant g5}

In order to fix the constant $g_5$, we follow the procedure described in \cite{nastase2015}. We start by defining the bulk-to-boundary field $\bar{v}(\omega,z)$ as the one that satisfies the condition
\begin{gather}
    v_\mu(\omega,z) = \bar{v}(\omega,z) v_\mu^0(\omega),
\shortintertext{where}
    \bar{v}(\omega,0) = 1.
    \label{eq: bulk-to-boundary condition}
\end{gather}
  Equation \ref{eq: bulk-to-boundary condition} is known as the bulk-to-boundary condition.

In the region of small values of $z$, the equation of motion \eqref{eq: eq motion} in terms of the bulk-to-boundary field takes the form
\begin{flalign}
    && \omega^2 \bar{v}(\omega,z) - \frac{1}{z} \bar{v}'(\omega,z) + \bar{v}''(\omega,z) = 0
    && \llap{(small $z$). \hspace{2em}}
\end{flalign}
The solution of this equation is
\begin{flalign}
    && \bar{v}(\omega,z) = 1 - \frac{1}{4} \omega^2 z^2 \ln(\omega^2 z^2)
    && \llap{(small $z$). \hspace{2em}}
    \label{eq: bulk-to-boundary field solution}
\end{flalign}

The two-point function in momentum space is calculated from
\begin{gather}
    \int d^4x e^{- \complexi p \cdot x} \average{J_\mu(x)J_\nu(0)},
\end{gather}
where $\average{J_\mu(x)J_\nu(0)}$ is given by
\begin{align}
    \average{J_\mu(x)J_\nu(0)}
    &= \frac{\delta^2 S_{\text{on shell}}}{\delta V^0_\mu(x) \delta V^0_\nu(0)},
\end{align}
where the on shell action is the action \eqref{eq: field action new metric} evaluated at the field that minimizes it. Using the fact that this field satisfies the equation of motion, it is possible to write the on shell action as
\begin{gather}
    S_{\text{on shell}} = -\frac{1}{2 g_5^2} \int_{z=0} d^4 x \sqrt{-g} e^{-\phi} g^{zz} g^{\mu\nu} V_\mu(x) \partial_z V_\nu(x).
\end{gather}

At the end, one can write the two-point function as
\begin{gather}
    \int d^4x e^{- \complexi p \cdot x} \average{J_\mu(x)J_\nu(0)} = (p_\mu p_\nu - p^2 g_{\mu\nu}) \Pi(p^2),
\shortintertext{with}
    \Pi(p^2) = -\lim_{z \rightarrow 0}{\frac{1}{g_5^2\, p^2}\, \frac{R}{z}\, e^{-\phi(z)}\, \bar{v}'(\omega,z)}.
\end{gather}

Using the bulk-to-boundary field obtained at \eqref{eq: bulk-to-boundary field solution}, we find
\begin{gather}
    \Pi(-  \omega^2) = -\frac{1}{2 g_5^2} R e^{-\phi(0)} \ln \omega^2,
    \label{eq: propagator - result}
\end{gather}
in the limit of large $\omega^2$.

The constant $g_5$ is fixed by imposing that the propagtor obained in \eqref{eq: propagator - result} matches the perturbative QCD result 
\begin{gather}
    \Pi(-  \omega^2) = -\frac{N_c}{24 \pi^2} \ln \omega^2,
\end{gather}
where $N_c = 3$ is the number of color charges. This gives
\begin{gather}
    g_5^{} = \sqrt{\frac{12 \pi^2}{N_c} R\, e^{-\phi(0)}} = 2 \pi \sqrt{R}\, e^{-\phi(0)/2}.
\end{gather}



\hspace{2\baselineskip}

\section*{Acknowledgements}

The authors are partially supported by CNPq --- Conselho Nacional de Desenvolvimento Científico e Tecnologico, by FAPERJ --- Fundação Carlos Chagas Filho de Amparo à Pesquisa do Estado do Rio de Janeiro and by  Coordenação de Aperfeiçoamento de Pessoal de Nível Superior --- Brasil (CAPES), Finance Code 001.


\end{document}